\documentclass[]{spie}  

\usepackage{amsmath,amsfonts,amssymb}
\usepackage{graphicx}
\usepackage[colorlinks=true, allcolors=blue]{hyperref}
\usepackage{caption}
\usepackage{subcaption}

\title{Estimating low-order aberrations through a Lyot coronagraph\\with a Zernike wavefront sensor for exoplanet imaging}

\author[a]{Raphaël Pourcelot}
\author[a]{Mamadou N'Diaye}
\author[b]{Greg Brady}
\author[a]{Marcel Carbillet}
\author[c]{Kjetil Dohlen}
\author[b]{Julia Fowler}
\author[b,c,f]{Iva Laginja}
\author[b]{Matthew Maclay}
\author[b]{James Noss}
\author[b]{Marshall Perrin}
\author[b,d]{Pete Petrone}
\author[e]{Emiel Por}
\author[c,f]{Jean-François Sauvage}
\author[b]{Rémi Soummer}
\author[c]{Arthur Vigan}
\author[b,g,h]{Scott Will}

\affil[a]{Université Côte d'Azur, Observatoire de la Côte d'Azur, CNRS, Laboratoire Lagrange, France}
\affil[b]{Space Telescope Science Institute, 3700 San Martin Drive, Baltimore, MD 21218, USA}
\affil[c]{Aix Marseille Univ., CNRS, CNES, LAM, Marseille, France}
\affil[d]{Sigma Space Corporation, 4600 Forbes Blvd, Lanham, MD 20706, USA}
\affil[e]{Leiden Observatory, Leiden University, P.O. Box 9513, 2300 RA Leiden, The Netherlands}
\affil[f]{DOTA, ONERA, Université Paris Saclay, F-92322 Châtillon, France}
\affil[g]{Institute of Optics, University of Rochester, Rochester, NY 14620, USA}
\affil[h]{NASA Goddard Space Flight Center, Greenbelt, MD 20771, USA}

\newcommand{\up}[1]{\textsuperscript{#1}}

\begin{document} 
\maketitle

\begin{abstract}
Imaging exo-Earths is an exciting but challenging task because of the 10\up{-10} contrast ratio between these planets and their host star at separations narrower than 100 mas. Large segmented aperture space telescopes enable the sensitivity needed to observe a large number of planets. Combined with coronagraphs with wavefront control, they present a promising avenue to generate a high-contrast region in the image of an observed star.

Another key aspect is the required stability in telescope pointing, focusing, and co-phasing of the segments of the telescope primary mirror for long-exposure observations of rocky planets for several hours to a few days. These wavefront errors should be stable down to a few tens of picometers RMS, requiring a permanent active correction of these errors during the observing sequence. To calibrate these pointing errors and other critical low-order aberrations, we propose a wavefront sensing path based on Zernike phase-contrast methods to analyze the starlight that is filtered out by the coronagraph at the telescope focus. In this work we present the analytical retrieval of the incoming low order aberrations in the starlight beam that is filtered out by an Apodized Pupil Lyot Coronagraph, one of the leading coronagraph types for starlight suppression. We implement this approach numerically for the active control of these aberrations and present an application with our first experimental results on the High-contrast imager for Complex Aperture Telescopes (HiCAT) testbed, the STScI testbed for Earth-twin observations with future large space observatories, such as LUVOIR and HabEx, two NASA flagship mission concepts.

\end{abstract}

\keywords{Zernike wavefront sensor, coronagraphy, high-contrast imaging, Low-order wavefront sensor.}

\section{INTRODUCTION}
\label{sec:intro}  

The 2020 Astrophysics Decadal survey is considering four large mission concepts, with two of them aiming to image Earth-like planets: the Large Ultraviolet Optical Infrared Surveyor (LUVOIR)\cite{LUVOIR2019} and the Habitable Exoplanet Imaging Mission (HabEx)\cite{Habex2019}. Relying on large segmented aperture telescopes, these missions will have the sensitivity and resolution to allow detection and spectral characterization of these planets with a contrast of 10\up{-10} at angular separations smaller than 100 mas from their host stars. 

Coronagraphy is one of the most promising solutions to suppress the light of an observed star. Several designs have been proposed over the past few years to reach these contrast levels with manufacturable solutions. We can cite the recent progress on the designs of the vortex phase mask coronagraph \cite{Ruane2018}, the Phase-Induced Amplitude Apodized Complex Mask Coronagraph \cite{Guyon2003}, or the Apodized Pupil Lyot Coronagraph (APLC) \cite{N'Diaye2016, Seo2019}. 

These coronagraphic solutions are sensitive to wavefront stability, a key aspect in exoplanet imaging, in particular for segmented aperture telescopes. There are several on-going studies dedicated to stability, e.g. the Ultra-Stable Telescope Research and Analysis (ULTRA) program \cite{Ultra2019}. In their initial report, the authors envision a subnanometric stability requirement for a LUVOIR-like 15m space telescope to observe exo-Earths around nearby stars. To correct for wavefront drifts during observations, these coronagraphs will be combined with active optics systems, which includes the use of deformable mirrors (DM). Wavefront sensing is also required to measure these small aberrations. The Zernike Wavefront Sensor (ZWFS) is a promising candidate for this application, thanks to its high sensitivity\cite{Guyon2005}. Recent laboratory results have demonstrated its capability for sensing aberrations at the picometric level \cite{Ruane2020}.

This sensor has also already been tested in preparation for the Nancy Grace Roman Space Telescope (NGRST, formerly known as WFIRST \cite{Shi2016}) with its coronagraphic instrument. The principle is to place the Zernike phase mask in the light filtered by the focal plane mask (FPM) of the coronagraph to measure low-order aberrations. In this configuration, an interaction matrix is used to calibrate the sensor response. 

In this contribution, our goal is to retrieve the wavefront errors analytically with a ZWFS and without interaction matrix by using a modified version of the sensor formalism \cite{N'Diaye2013, Wallace2011, Ruane2020}. To validate this approach, we use the High-Contrast imager for Complex Aperture Telescopes (HiCAT) testbed. Located at the Space Telescope Science Institute (STScI) in Baltimore, this bench aims to advance high-contrast technologies for Earth-like observations with future large segmented aperture telescopes \cite{hicat1, hicat2, hicat3, hicat4, hicat5, hicat6}.

In this paper, we describe the new formalism and algorithms for the wavefront calibration in the presence of a Classical Lyot Coronagraph (CLC). We detail the numerical simulations to validate our approach, using the HiCAT testbed emulator that includes an optical simulation model. Finally we will present preliminary experimental results on the testbed.


\section{FORMALISM}
\label{sec:formalism}

\subsection{Testbed setup presentation} \label{sec:setup}

\begin{figure}
    \centering
    \includegraphics[width=\columnwidth]{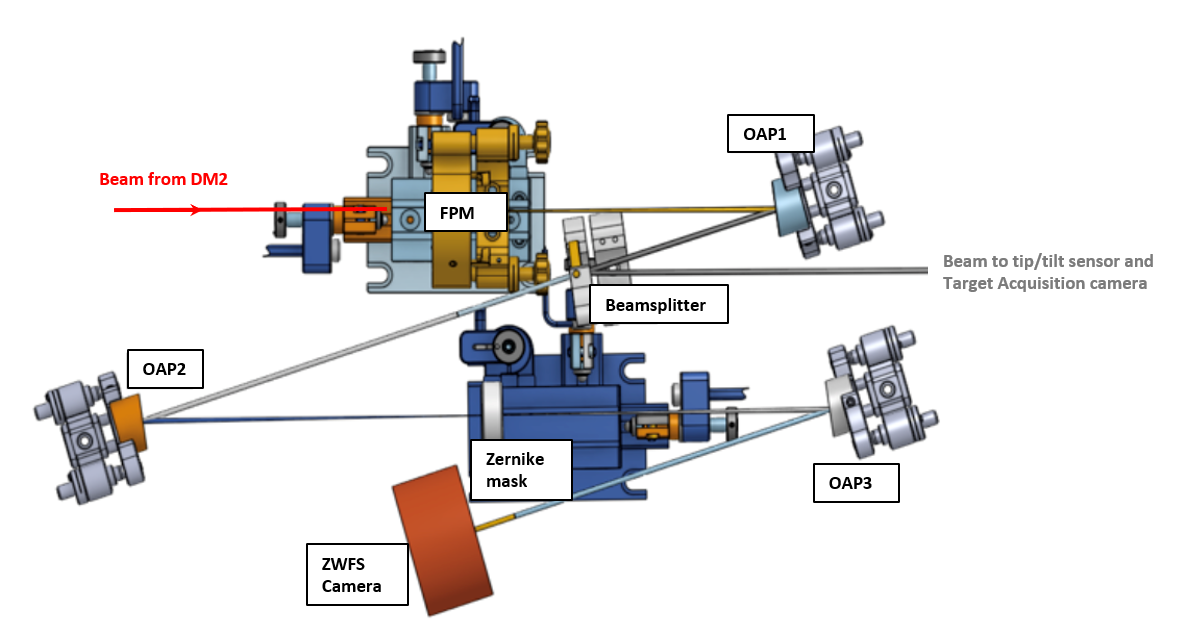}
    \caption{3D model of the implementation of the ZWFS on HiCAT. The FPM and the Zernike mask are located in two focal planes. There are also two pupil planes, one between the beamsplitter and OAP2, and one at the ZWFS camera.}
    \label{fig:hicat_blueprint}
\end{figure}

\begin{figure}
    \centering
    \includegraphics[width=\columnwidth]{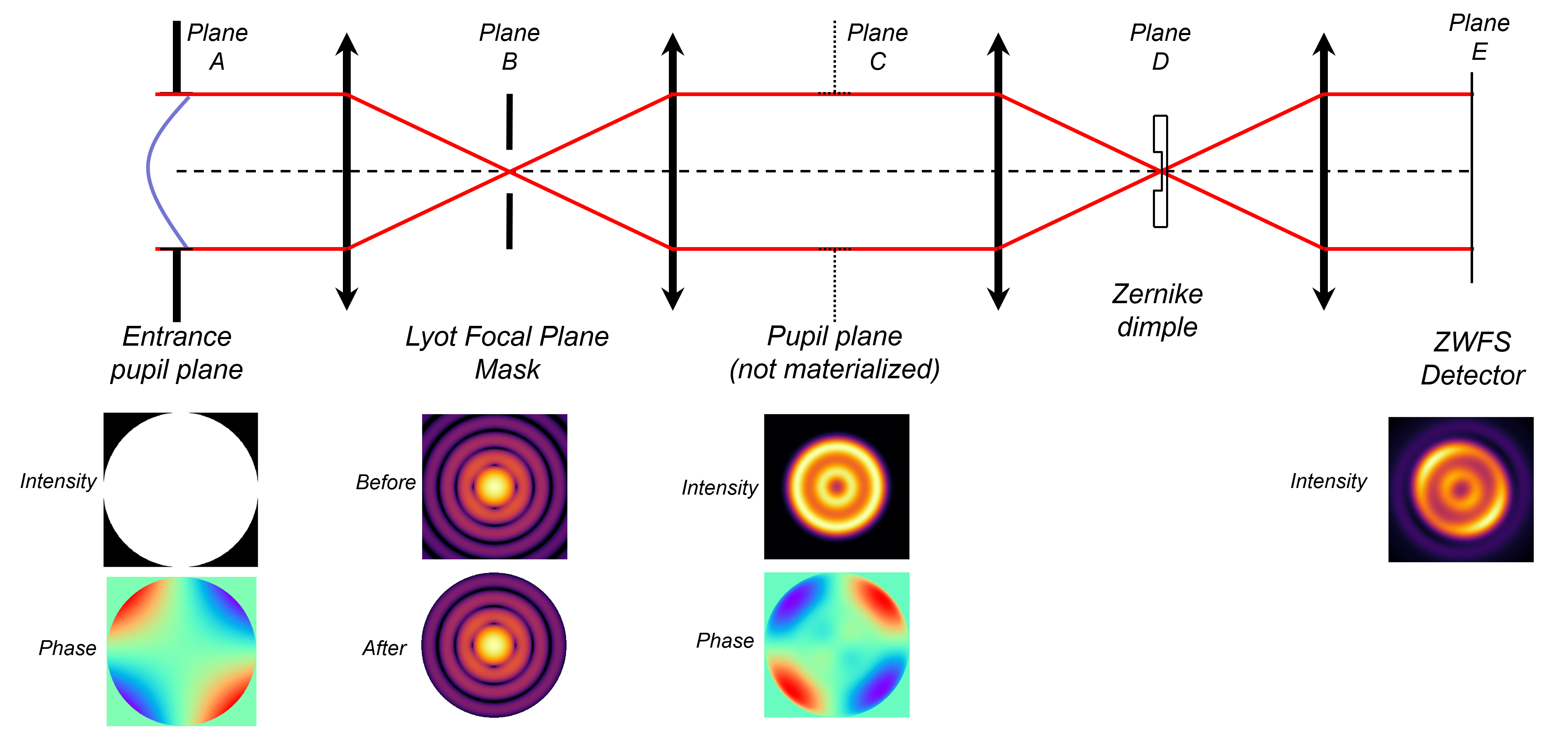}
    \caption{Simplified optical layout of the ZWFS on HiCAT. The pupil plane A is conjugated with DM1. The FPM is located in plane B. There is an intermediate relay pupil in plane C. The Zernike mask is in the focal plane D and the detector is located in the re-imaged pupil plane E. Below the scheme, the images represent the intensity or the phase of the electric field in the different planes.}
    \label{fig:blueprint}
\end{figure}

Here we analyze the case of a CLC with reflective optics. The implementation of the ZWFS on HiCAT is presented in Fig.~\ref{fig:hicat_blueprint}. In this configuration, the focal plane mask (FPM) is a reflective mirror with a pinhole centered on the source image. The FPM diameter is $8.556 \lambda/D_{pup}$ at a wavelength $\lambda$ of 640 nm, $D_{pup}$ denoting the diameter of the entrance pupil. Picking up this beam is then straightforward, by placing off-axis parabolas (OAP) to re-image the focal plane in which the Zernike phase mask is placed. This mask consists of a circular dimple machined into the front face of a fused silica substrate by the aid of photolithographic reactive ion etching. It has an angular diameter of $1.02\,\lambda/D_{pup}$. The Zernike sensor will produce interference between a reference wave created by the mask and the phase errors present in the system, converting them into intensity variations on the camera in pupil plane E.
The bench uses two DMs, one in pupil plane (DM1) and the other out-of-pupil plane (DM2) upstream of the FPM. 
Fig.~\ref{fig:blueprint} displays a scheme of this setup in which the mirrors have been replaced by lenses for the sake of clarity.

The beam going through the FPM pinhole is filtered with a cutoff spatial frequency of 4.278\,cycles/pupil (c/p) on DM1. This will have an effect on both the amplitude and phase of the electric field that will be measured by the Zernike mask. The amplitude of the clear pupil in the entrance pupil plane A will be modified, leading to a filtered amplitude in pupil plane C. The amplitude becomes heterogeneous with variations inside and outside the geometric pupil. 

To describe the filtered wave, we use the formalism showed in N'Diaye et al.\cite{N'Diaye2013}. We denote $\psi_A$ the complex amplitude in pupil plane A. It can be expressed with the pupil amplitude $P$ and phase $\varphi$ as
\begin{equation}
    \psi_A = P e^{i\varphi}.
\end{equation}

The transmission of the FPM is a top-hat function $M_c$, whose value is 1 within the mask and 0 outside. 
The incoming wave on the FPM $\psi_{B}$ is obtained with the Fourier transform of $\psi_A$, noted $\widehat{\psi_A}$
\begin{equation}
    \psi_B = M_c\widehat{\psi_A}.
\end{equation}

In the following pupil plane C, the complex amplitude $\psi_C$ can be written as
\begin{align}
    \psi_C &= \widehat{\psi_B} \\
           &= \widehat{M_c} \otimes \psi_A,
\end{align}
in which $\otimes$ denotes the convolution product. 

Without any filtering operation in plane C, the electric fields in planes B and D before Zernike mask application are the same. We define $t$ the transmission of the Zernike mask which can be written as
\begin{equation}
    t = 1 - \left(1-e^{i\theta}\right)M, 
\end{equation}

with M the top-hat function defining the phase mask dimple and $\theta$ the phase shift introduced by the mask.

The wave $\psi_D$ after the Zernike mask writes as
\begin{align}
    \psi_D &= t\psi_B.
\end{align}

Finally, the amplitude $\psi_E$ in the detector pupil plane can be written as
\begin{align}
    \psi_E &= \widehat{\psi_D} \\
           &= \psi_C - (1-e^{i\theta})\widehat{(M\cdot M_C})\otimes \psi_A.
\end{align}

The Zernike mask dimple diameter is $1.02\,\lambda/D_{pup}$. As this mask is smaller than the FPM pinhole, we have $M\times M_C = M$.
Using the definition of the reference wave $b = \widehat{M}\otimes \psi_A$ in the literature\cite{N'Diaye2013}, the reference wave is unchanged by the filtering of the FPM.

We thus obtain
\begin{align} \label{eq:final_amp}
    \psi_E = \psi_C - (1-e^{i\theta})b.
\end{align}

The ZWFS here works in the same way as the traditional Zernike sensor. The only difference lies in the fact that the sensor here is measuring the FPM-filtered version $\psi_C$ instead of the original field $\psi_A$. The formalism given in the literature \cite{N'Diaye2013, Vigan2018} is therefore still valid to reconstruct the electric field $\psi_C = \widehat{M_c}\otimes\psi_A$.

\subsection{Analytical phase reconstruction}

To perform analytical phase measurements with the ZWFS, we use the open-source code pyZELDA \cite{Vigan2018} that implements the formalism from N'Diaye et al \cite{N'Diaye2013}. This library has been used in particular on a ground-based telescope, on the exoplanet imager VLT/SPHERE \cite{N'Diaye2016, Vigan2019}. The phase and intensity are related through a general equation. The code implements an approximation of this equation based on the second-order Taylor expansion for $\varphi$ in the regime of small aberrations. This yields a second order polynomial equation between the intensity on the detector $I_E$ and the phase $\varphi_C = M_c \otimes \varphi$ in pupil plane C, as follows:

\begin{equation} \label{eq:2ndorder}
    I_E = P_C^2 + 2\,b^2(1-\cos{\theta}) + 2\,P_C\,b\left[\varphi_C \sin{\theta} - (1-\varphi_C^2/2)(1-\cos{\theta})\right],
\end{equation}
where $P_C$ denotes the modulus of $\psi_C$ which can be rewritten as $P_0(1-\epsilon)$, with $P_0$ the perfect pupil amplitude and $\epsilon$ the small zero-mean amplitude errors. 

Originally, pyZELDA uses the assumption that the pupil is homogeneous with binary values to simplify this equation. This is equivalent to consider $P_C=0$ or $P_C=1$ in Eq.~\ref{eq:2ndorder}. Such an assumption is valid for many pupil designs with features such as monolithic apertures, secondary mirror obscuration, spiders, segmented primary mirrors or shaped pupil apodizers. However, the filtered pupil amplitude $\psi_C$ does not fall within the assumption of $\epsilon << 1$ which means the $P_C=0$ or $P_C=1$ assumption cannot be used here.

Considering large pupil amplitude variations, such as the filtered wave amplitude $\psi_C$ here or a given grey-value apodization, we use eq.~\ref{eq:2ndorder} by keeping $P_C$. We derive the full solution, where the discriminant $\Delta$ is given by
\begin{align}
    \Delta = \sin^2{\theta} - \frac{2 \cdot (b-P_C)\cdot (1-\cos^2{\theta})}{P_C}  - \frac{P_C^2-I_E\cdot(1-\cos{\theta})}{P_C\cdot b} .
\end{align}

The solution then writes as
\begin{align}
    \varphi_C = \frac{-\sin{\theta}+\sqrt{\Delta}}{1-\cos{\theta}}.
\end{align}

To estimate $\varphi_C$, we need to know $P_C$ and $b$. Both can be estimated from the clear pupil intensity $I_C = P_C^2$ that is acquired with a measurement without the Zernike mask. Experimentally, this is performed by moving the mask a few millimeters away from the optical axis. For the configuration without FPM, we can retrieve $P_C$ and $b$ with $P_C=\sqrt{I_{C}}$ as $P_C$ is positive and $b=\widehat{M}\otimes P_C$. For the configuration with FPM, using the wavefront reconstruction of Eq.~\ref{eq:2ndorder} will lead to discontinuities in $\varphi_C$ with $\pi$ shifts. To understand these discontinuities, we simulate $\psi_C$ in the absence of aberrations and note that this term is equal to $\pm P_C$. To account for this effect in the CLC case, we introduce these sign changes in the expression of $b=\widehat{M}\otimes P_C$.

\section{Zernike WFS performance}
\label{sec:simulations}

\subsection{ZWFS measurement procedure}

HiCAT comes with a realistic simulator controlled in the same way as the hardware\cite{hicat6}: a script running on the simulator will run on the bench. We use this powerful tool to test our wavefront reconstruction algorithm and predict the ZWFS performance. The operations consist in introducing phase aberrations on one of the DMs and analyze open-loop measurements and closed-loop corrections. 

Since there are residual aberrations on the bench, we use differential phase measurements: we first perform a reference phase measurement with flattened DMs. We then introduce the aberration on DM1 or DM2 on top of the flat map, perform a phase measurement with the ZWFS and subtract the reference phase. Absolute measurement could be performed, but a precise calibration of the non-common path aberrations 
existing between the ZWFS path and the science path on HiCAT is required. This will be part of a future work. 

\subsection{Open-loop simulations} \label{sec:openloop}
\begin{figure}
    \centering
    \includegraphics[width=\columnwidth]{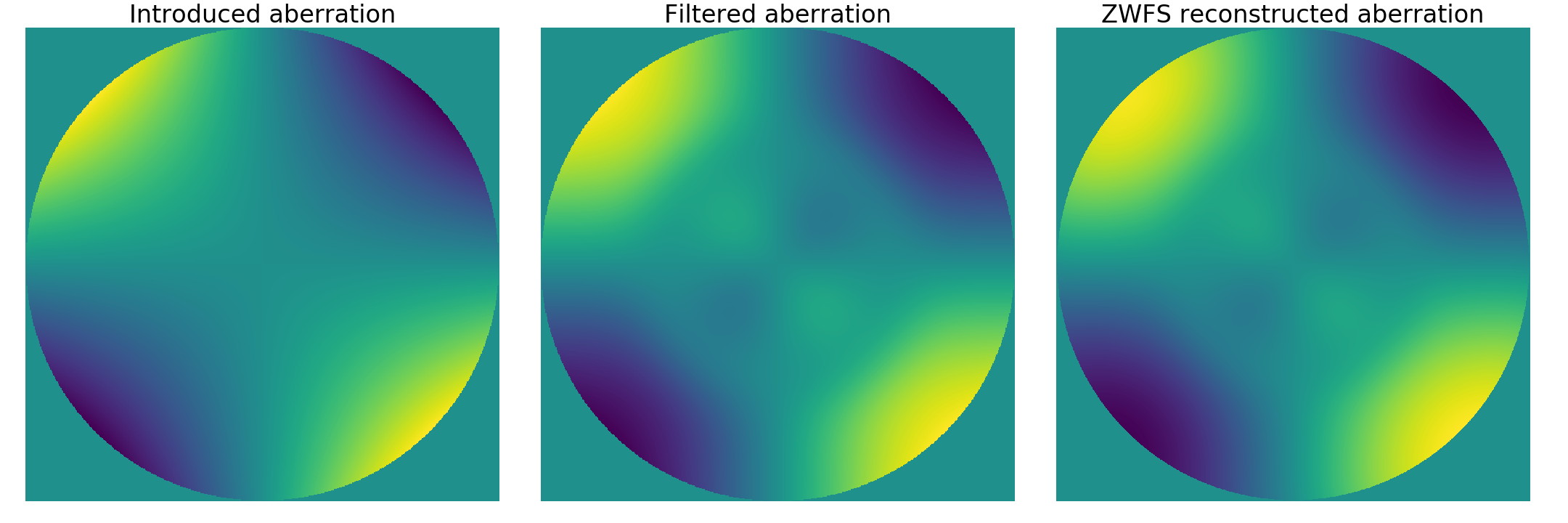}
    \caption{Comparison of the introduced astigmatism on DM1 (left), the theoretical filtered mode in plane C (middle) and the reconstructed aberration in simulation (right).}
    \label{fig:comp_astig}
\end{figure}

We assess the ZWFS performance in open loop by comparing the introduced aberrations and the corresponding measurements. The first test consists in introducing known aberrations such as Zernike modes. Because of the spatial filtering induced by the FPM, the Zernike modes appear filtered in pupil plane C. Fig.~\ref{fig:comp_astig} shows a comparison between the introduced aberration (left) and the theoretical filtered aberration (middle), illustrating the impact of the FPM filtering on the Zernike mode.

\begin{figure}
    \centering
    \includegraphics[width=.7\columnwidth]{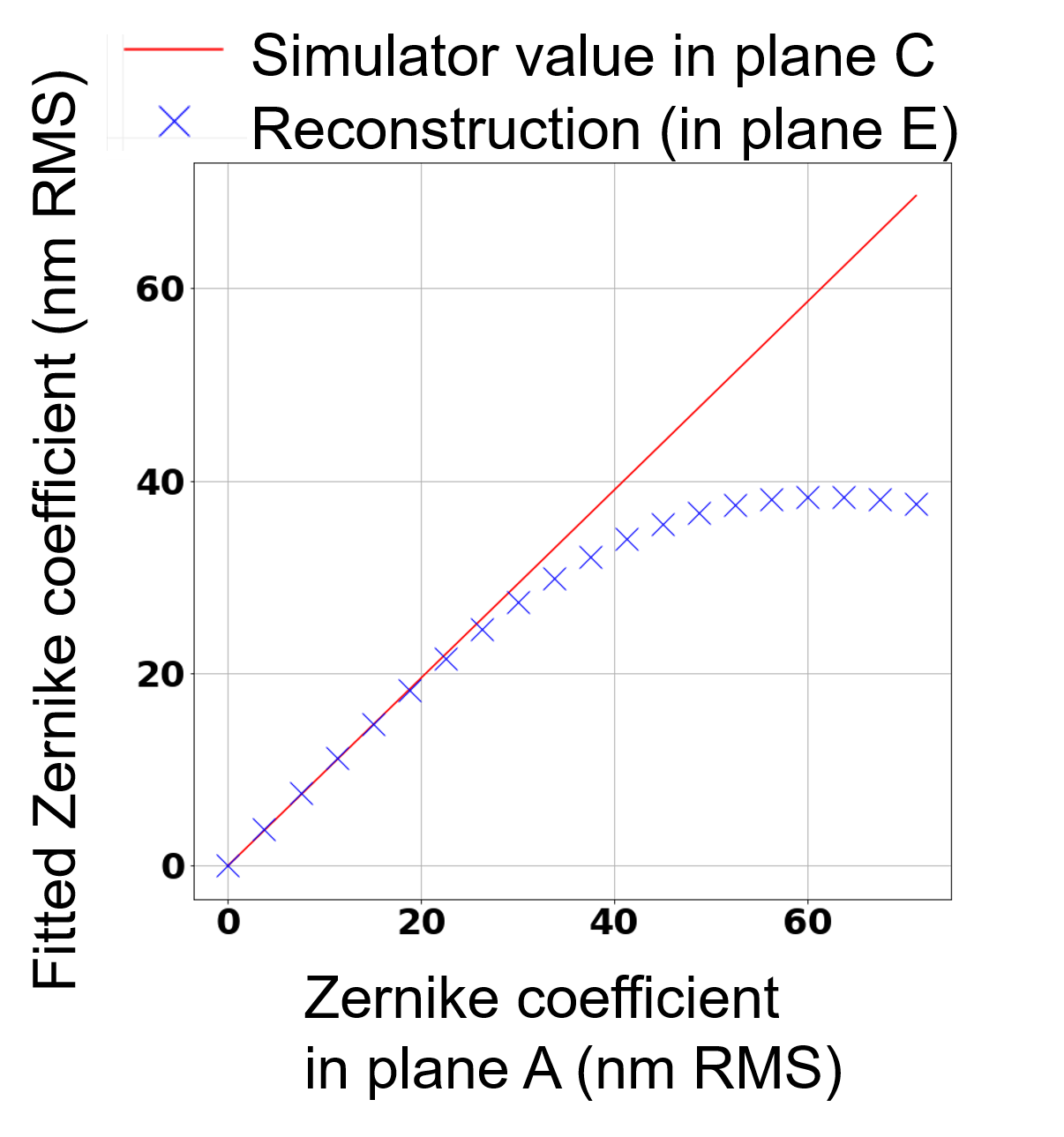}
    \caption{Fitting of the Z4 astigmatism in the theoretical wavefront and in the reconstructed wavefront by the ZWFS as a function of the introduced Zernike coefficient, in simulation.}
    \label{fig:zernike_sensitivity}
\end{figure}

Fig~\ref{fig:comp_astig} right image displays the reconstructed aberration in simulation, showing a good agreement with the theoretical filtered mode. The algorithm manages to retrieve the introduced aberration for a typical value of a few nm RMS. The sensor shows a similar response to the standard ZWFS implementation without FPM filtering. Fig.~\ref{fig:zernike_sensitivity} shows the sensor response. The behavior is linear in the small phase regime and then underestimates the introduced aberration. This is inherent to the linearity range of the ZWFS \cite{N'Diaye2013, N'Diaye2016}.


\begin{figure}
    \centering
    \includegraphics[width=.7\columnwidth]{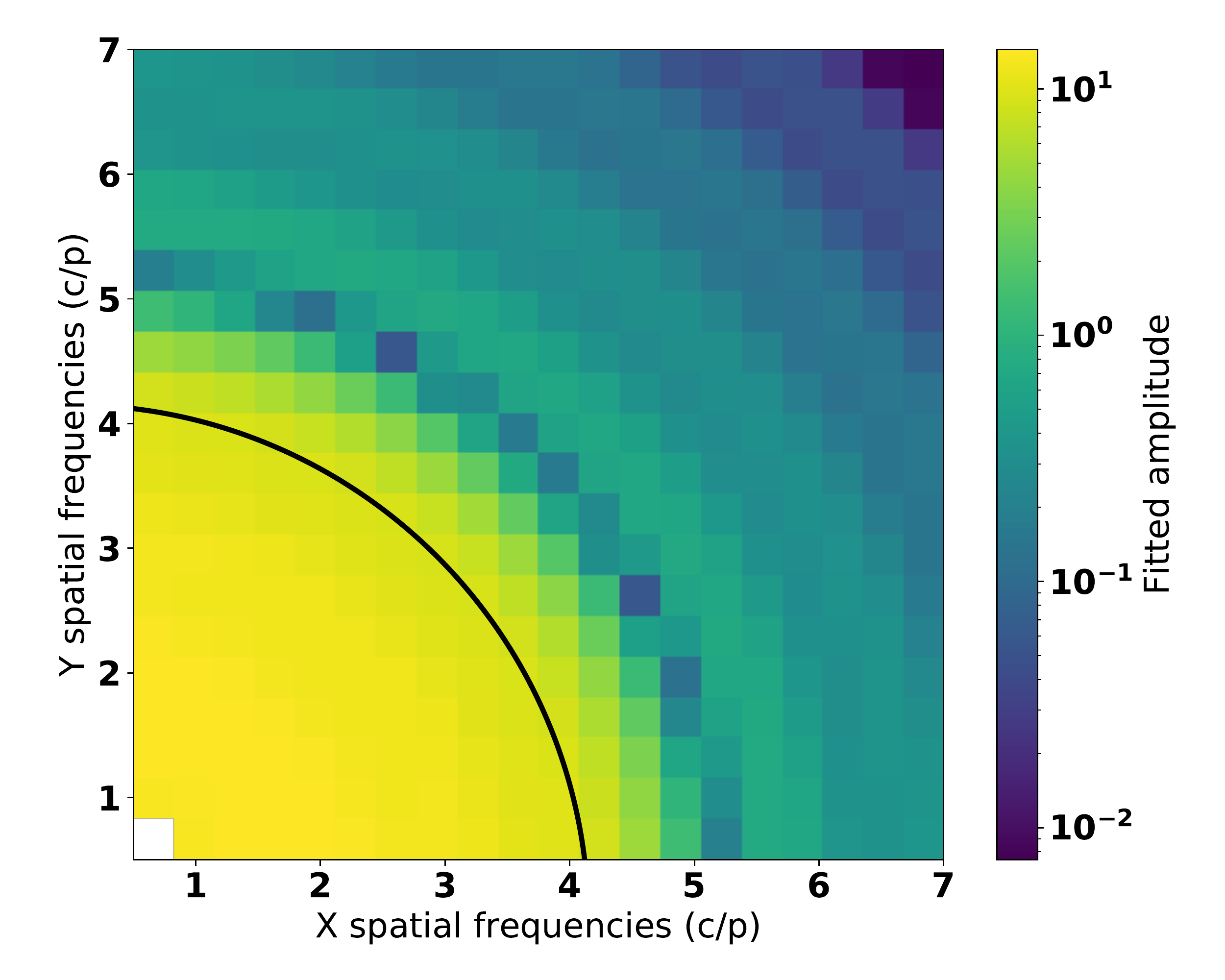}
    \caption{2D response of the "FPM + ZWFS" system. Each pixel intensity corresponds to the fitted amplitude of the introduced sinus wave in X and Y directions. The black curve corresponds to the theoretical cut-off of the FPM, at 4.278 c/p.}
    \label{fig:fourier_2D}
\end{figure}

Describing the filtered aberrations is a challenging task as filtered Zernike modes are not a basis of the filtered phases, which means that filtering introduces mode coupling. For example, astigmatism generates higher-order astigmatism, coma generates tip, tilt and 5\up{th}-order coma. To avoid this effect, we use an alternative basis, the Fourier modes: it is still a basis of the space of the phases, even after the Fourier filtering due to the FPM. To assess the response of the "filter + sensor" system, we introduce sinusoidal shapes with various spatial frequencies on DM1 with an amplitude of 10 nm. We then fit the corresponding Fourier mode in the wavefront reconstructed by the ZWFS. Fig.~\ref{fig:fourier_2D} shows the response of this test that scans spatial frequencies in both X and Y directions, from 0.5\,c/p to 7\,c/p. It shows a clear cutoff at the spatial frequency corresponding to the FPM radius. The residual noise can be attributed to the fact that the Fourier modes are not a well-defined basis on a circular pupil. Therefore, the induced spots in focal plane have the same diffraction pattern as the PSF of an on-axis point source, here an Airy pattern. This means that when the focal plane is cropped by the FPM, the Airy pattern of the spots might be partially cropped too, inducing some side effect and mode coupling.

\subsection{Close-loop simulations}

Following the reconstruction performance of the setup in open loop simulations, a key aspect of this sensor is to control the low-order aberrations during a high-contrast observing run. Such a run could start with the generation of a dark hole (DH) in the coronagraphic image of an observed star, an area in which starlight has been removed with wavefront control algorithms using two DMs \cite{Malbet1995, Give'on2007, Give'on2009}. HiCAT uses stroke minimization as DH algorithm to determine the optimal DM shapes \cite{Pueyo2009}. In our closed loop simulation, the goal is to estimate the quality of the correction for an aberration appearing during the observations. We use the following closed-loop procedure to mimic this scenario: we introduce an aberration on top of DM1 DH shape, here 40 nm RMS of astigmatism; we perform a phase measurement with ZWFS; finally, we apply the opposite of the measured phase map on DM1 with a gain of 0.5 and reproduce measurement and correction for several iterations. 

\begin{figure}
    \centering
    \includegraphics[width=\columnwidth]{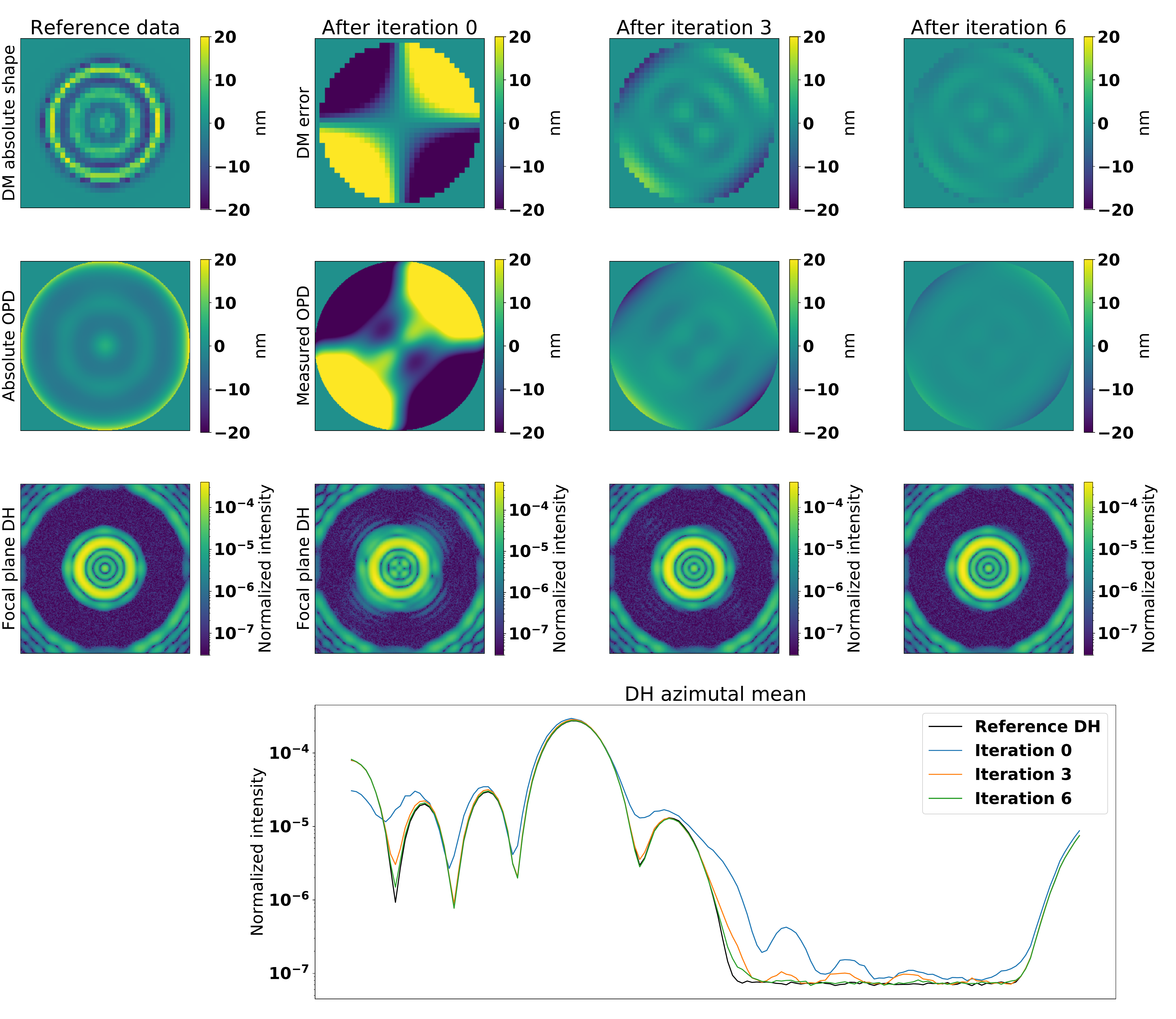}
    \caption{Simulation of dark-hole maintenance with a ZWFS on the low spatial frequencies. First column presents the absolute status of the bench before introducing an aberration on DM1. The following columns represent iterations 0, 3 and 6 from left to right. Iteration 0 corresponds to the introduced aberration without any correction. \textbf{First row}: difference of the current DM1 surface and the DH pre-computed shape. \textbf{Second row}: phase measured by the ZWFS after removal of the phase due to the DH DM shapes. \textbf{Third row}: Corresponding coronagraphic image of the source with DH on the HiCAT science path. \textbf{Fourth row}: Azimutally averaged intensity profiles of the coronagraphic images that are displayed in the third row for different iterations.}
    \label{fig:close_loop}
\end{figure}

The result of this simulation is shown in Fig.~\ref{fig:close_loop}. After 6 iterations, the initial DH contrast is almost recovered as shown in the bottom row of frames and the plot. In the final iterations, the residual shape of the DM (top row) is the higher-order component left from the introduced aberration. The low-order aberration seen by the ZWFS tends to 0 (see middle row), showing that all the aberrations after FPM filtering are completely corrected. The contrast recovery needs slightly more iterations at shorter than longer separations from the inner working angle of the DH.

\subsection{Preliminary experimental results on HiCAT}

\begin{figure}
    \centering
    \includegraphics[width=\columnwidth]{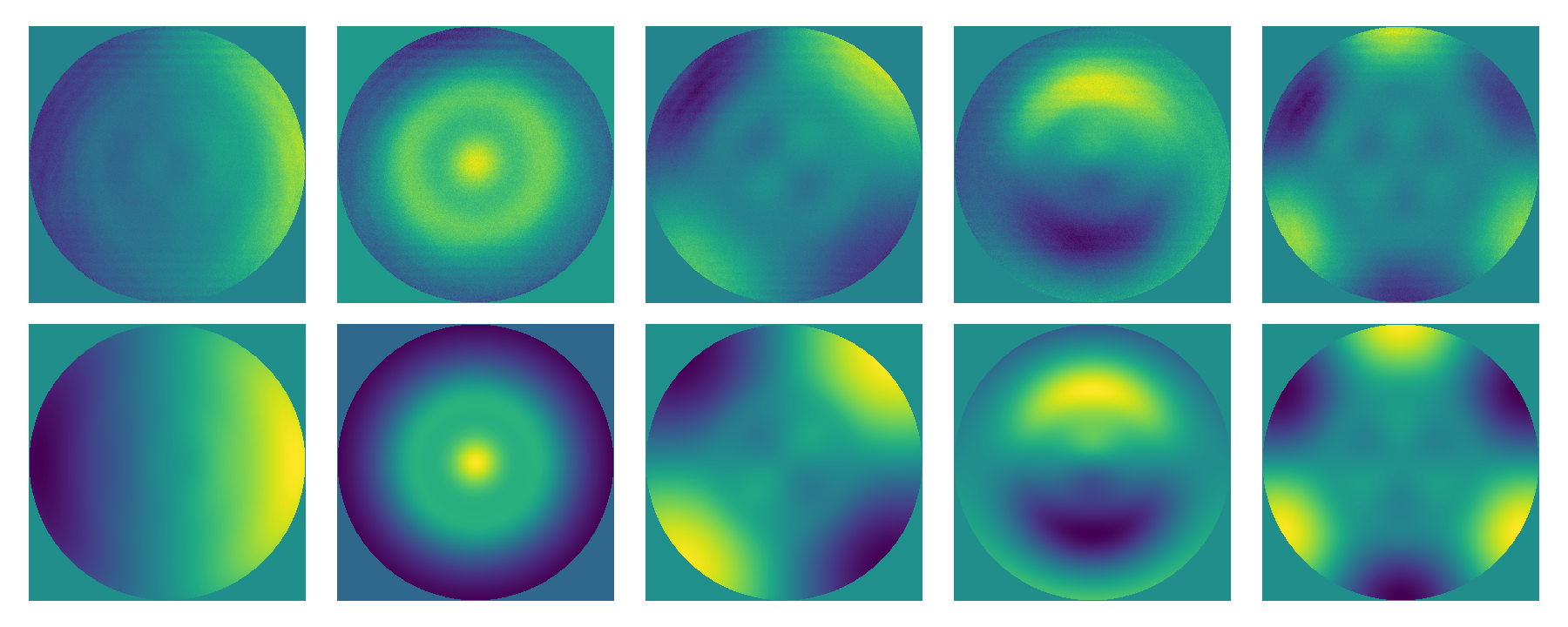}
    \caption{Qualitative comparison of the reconstructed modes in simulation (top row) and experiment (bottom row). Each column presents an introduced Zernike mode on DM2. From left to right: tip, defocus, astigmatism, coma, trefoil.}
    \label{fig:comp_sim_exp}
\end{figure}

The recent installation and alignment of a ZWFS on HiCAT enabled tests to validate the performed simulations on hardware. We perform the same measurements as in Sec.~\ref{sec:openloop} by introducing Zernike modes. We first flatten the DM with a pre-existing calibrated flat map, measure the residual aberrations, and introduce Zernike modes on DM2. As this mirror is located in an out-of-pupil plane, phase on this device will convert into amplitude errors due to Fresnel propagation effects. On HiCAT, the Talbot length for the low-order aberrations considered is much larger than the distance between the two DMs. As a result, there will not be  any noticeable amplitude and phase coupling on the ZWFS when introducing phase patterns on DM2. We perform the test with the first ten Zernike modes. Fig.~\ref{fig:comp_sim_exp} displays the results for the simulator and the in-lab experiment, qualitatively showing good agreement between them. These promising results pave the way for more advanced tests with quantitative analysis. 
    
\section{CONCLUSION}
\label{sec:conclusion}

In this work, we addressed phase reconstruction with a ZWFS in a beam filtered by the FPM of a Lyot coronagraph. The analytical formalism of this sensor for phase reconstruction was adapted to any kind of pupil amplitude aberration, including non-uniform amplitudes due to FPM filtering. Using a modified version of the pyZELDA code as well as the HiCAT simulator and bench, we showed successful examples of phase reconstruction. With the HiCAT simulator, we introduced aberrations on DM1, such as Zernike or Fourier modes and compared them with the measurements, showing good agreement between theory and simulations. In addition, we showed the closed-loop correction with ZWFS for an aberration introduced after generation of a DH, in simulation.

Finally, we presented laboratory measurements on HiCAT with Zernike modes on DM2. Our preliminary results showed a good agreement between simulation and experiment, proving encouraging for the validation of our implementation. This work paves the way for further tests such as closed loop control on HiCAT, stability measurement, and fine tip/tilt compensation. The current framework used a circular pupil and a CLC in monochromatic light. This can be extended to other configurations including a segmented pupil, an APLC, and work in broadband light. Such aspects will be addressed in the forthcoming months to advance the control of low-order aberrations for exoplanet imaging with large segmented aperture telescopes. 

\acknowledgments 
R.P. acknowledges PhD scholarship funding from R\'egion Provence-Alpes-C\^ote d'Azur and Thales Alenia Space. This work was also supported in part by the National Aeronautics and Space Administration under Grant 80NSSC19K0120 issued through the Strategic Astrophysics Technology/Technology Demonstration for Exoplanet Missions Program (SAT-TDEM; PI: R. Soummer)

\bibliography{report} 
\bibliographystyle{spiebib} 

\end{document}